\documentclass[english,a4paper]{article}
\usepackage{dcolumn}
\usepackage{bm}
\usepackage{graphicx}
\usepackage{amsmath}
\usepackage{amsfonts}
\usepackage{amssymb}
\usepackage{amsthm}
\usepackage{amstext}
\usepackage{amsbsy}
\usepackage{amsopn}
\usepackage{amscd}
\usepackage{amsxtra}

\def\openone{\leavevmode\hbox{\small1\kern-3.3pt\normalsize1}}

\usepackage[usenames,dvipsnames]{color}

\begin{document}
\title{Purity Speed Limit of Open Quantum Systems from Magic Subspaces}
\author{A. A. D\`iaz V., V. Martikyan, S. J. Glaser\footnote{Department of Chemistry, Technical University of Munich, Lichtenbergstrasse 4, 85747 Garching, Germany}, D. Sugny\footnote{Laboratoire Interdisciplinaire Carnot de
Bourgogne (ICB), UMR 6303 CNRS-Universit\'e Bourgogne-Franche Comt\'e, 9 Av. A.
Savary, BP 47 870, F-21078 Dijon Cedex, France, dominique.sugny@u-bourgogne.fr}}

\maketitle

\begin{abstract}
We introduce the concept of Magic Subspaces for the control of dissipative $N$- level quantum systems whose dynamics are governed by Lindblad equation. For a given purity, these subspaces can be defined as the set of density matrices for which the rate of purity change is maximum or minimum. Adding fictitious control fields to the system so that two density operators with the same purity can be connected in a very short time, we show that magic subspaces allow to derive a purity speed limit, which only depends on the relaxation rates. We emphasize the superiority of this limit with respect to established bounds and its tightness in the case of a two-level dissipative quantum system. The link between the speed limit and the corresponding time-optimal solution is discussed in the framework of this study. Explicit examples are described for two- and three- level quantum systems.
\end{abstract}

\section{Introduction}
Controlling quantum dynamics to achieve a specific task in minimum time is a crucial prerequisite in many fields extending from quantum technologies and quantum optics to magnetic resonance and molecular physics~\cite{glaser15,brif,RMP,dong,alessandrobook}. This problem can be solved by using tools of optimal control theory (OCT)~\cite{pont}. However, deriving a rigorous optimal solution is a highly non trivial task which can only be done in low dimensional closed or open quantum systems~(see~\cite{alessandro,boscain,hegerfeldt,garon,khaneja1,khaneja2,lapert,bonnard} to mention a few). Different numerical optimization methods have been developed to approximate the time-optimal trajectory~\cite{grape,reich,gross,doria,calarco}. The many local minima of the control landscape make it very difficult to find a good approximation and lead generally to an upper bound of the minimum time. On the other side, lower bounds on the time can be established in the framework of quantum speed limits (QSL)~\cite{QSLreview,QSLreview2} where the time is expressed as a ratio between the distance to the target state and the dynamical speed of evolution. This approach has been the subject of an intense development in recent years with applications in quantum computing~\cite{lloyd,giovannetti}, quantum metrology~\cite{giovannetti2,alipour,chin,demkowicz} and quantum thermodynamics~\cite{demkowicz,deffnerthermo,qthermo}. Speed limits have been also introduced in classical systems, showing that this concept is not limited to quantum dynamics~\cite{shanadan,okuyama}. The tightest of these bounds is generally difficult to estimate~\cite{campaioli,pires} and very few connections exist with optimal control protocols~\cite{hegerfeldt,calarco,bason}. First established for closed quantum systems on the basis of Heisenberg time-energy uncertainty relation, QSL have been recently extended to open systems in Markovian and non-Markovian regimes~\cite{lidar,deffner,taddei,campo,kosloff,campaioli2,funo,hutter,brody}. In this setting, different QSL have been proposed according to the target state to reach by the quantum system~\cite{QSLreview}. In particular, bounds are known for a specific final density operator~\cite{deffner,campo}, but also for the rate of entropy or purity evolution~\cite{hutter,kosloff,brody,funo}. In this study, we consider the Purity Speed Limit (PSL) established in~\cite{kosloff} for systems coupled to a Markovian environment as a reference for the minimum time of purity evolution. The bounds of Ref.~\cite{kosloff} are said to be cumulative in the sense that they do not describe the instantaneous variation rate, but the global dynamics of the purity between the initial and final states. A key advantage of this point of view is the fact that this limit can be determined directly from relaxation parameters without computing the dynamics of the density operator.

This paper explores the time-optimal control of purity evolution in dissipative quantum systems whose dynamics are governed by Lindblad equation. Many studies have explored the control of these open quantum systems. Controllability results have been established and the set of reachable states can be characterized~\cite{altafini1,altafini2,dive,schulte1,schulte2}. Numerical optimal control procedures have been applied with success (see the recent review \cite{kochreview} and references therein). Geometric or analytic optimal control results can be achieved in low-dimensional open quantum systems. The time-optimal control of a two-level system has been solved in a series of papers~\cite{lapert,lapert:2013,mukherjee,tannor,bonnard1,bonnard2}, showing the key role of geometric objects, namely the magic plane and axis~\cite{lapert:2013} in the derivation of the optimal control process. In the Bloch representation, the magic plane is parallel to the equatorial plane and is defined as the set of points for which the shrinking of the purity of the density operator is maximum. The magic axis is the axis corresponding to diagonal density matrices. The generalization of this approach to higher dimensional quantum systems is difficult and much more involved from a mathematical point of view. Some results have been established in the optimal cooling process of three-level quantum systems~\cite{khaneja3}. A difficulty of the control problem comes from the fact that all the density matrices of a given purity cannot be connected by unitary dynamics generated by the control fields~\cite{schirmer}. Relaxing this constraint by adding fictitious control terms, we show in this study that the time-optimal control of the purity evolution can be solved. To this aim, we introduce the magic subspaces, which are higher-dimensional generalizations of the magic plane and axis. For a given purity of the density operator, the magic subspaces can be defined as the set of density matrices for which the rate of purity change is maximum or minimum. They can be viewed as the counterpart of decoherence-free subspaces~\cite{DFS}, which are defined as the subspaces with no decoherence, and thus a constant purity. The addition of non-physical control parameters leads only to a lower bound of the original control time. In other words, this approach can also be interpreted as a new way to derive PSL. This limit is tight for a two-level quantum system and corresponds exactly to the time-optimal solution. In a three-level quantum system, the minimum time is estimated by using numerical optimization techniques~\cite{grape}. We show that the speed limit time gives a good approximation of this minimum time. In the general case, we highlight the efficiency of this method by comparing this new bound to the speed limits derived in~\cite{kosloff}. We provide a simple asymptotic expression of PSL when the dephasing rate goes to infinity. Explicit computations are presented for a three-level quantum system.

The paper is organized as follows. The model system and the general approach for a $N$- level quantum system are described in Sec.~\ref{sec2} and \ref{sec2new}. Sections~\ref{sec3} and~\ref{sec4} focus on two specific examples in two and three-level quantum systems, respectively. A comparison with the existing PSL and numerical optimal computations is made in Sec.~\ref{sec5}. Conclusion and prospective views are given in Sec.~\ref{sec6}. Technical computations are reported in the Appendices. PSL of~\cite{kosloff} are briefly recalled in Appendix~\ref{secappa}. The computation of these limits for two- and three- level quantum systems is discussed. The dynamics and the PSL of dissipative three-level quantum systems are respectively described in Appendices~\ref{secappb} and~\ref{secappc}.
\section{The model system}\label{sec2}
We consider a dissipative $N$-level quantum system whose dynamics are governed by Lindblad equation~\cite{breuerbook}. The system is described by a density operator
$\rho(t)$ which is a positive Hermitian operator acting on a Hilbert space $\mathcal{H}$ spanned by the canonical orthonormal basis $\{|k\rangle\}_{k=1,N}$ of the field-free Hamiltonian $H_0$. The evolution equation can be written in atomic units (with $\hbar=1$) as:
\begin{equation}\label{eq1}
i\dot{\rho}=[H_0+H_I,\rho]+\mathcal{L}_D(\rho),
\end{equation}
where the unitary and dissipative parts of the equation are represented respectively by the Hamiltonian $H=H_0+H_I$ and the operator $\mathcal{L}_D$. In the Lindblad equation, $\mathcal{L}_D$~\cite{lindblad,gorini} can be expressed as:
\begin{equation}\label{eq2}
\mathcal{L}_D(\rho)=\frac{1}{2}\sum_{l,m=1}^{N^2-1}a_{lm}([V_k\rho,V_{k'}^\dagger]+[V_k,\rho V_{k'}^\dagger]
\end{equation}
where the operators $V_k$ are trace-zero and orthonormal, $\textrm{Tr}(V_{k'}^\dagger V_k)=\delta_{k'k}$. A canonical choice is given by the generalized Pauli matrices:
\begin{equation}\label{eqpauli}
\begin{cases}
\sigma_{m,n}^x=\frac{1}{\sqrt{2}}(|m\rangle\langle n|+|n\rangle\langle m|)\\
\sigma_{m,n}^y=\frac{i}{\sqrt{2}}(|m\rangle\langle n|-|n\rangle\langle m|)\\
\sigma_{m,n}^z=\frac{1}{\sqrt{m+m^2}}(\sum_{k=1}^m|k\rangle\langle k|-m|m+1\rangle\langle m+1|)
\end{cases}
\end{equation}
with $1\leq m\leq N-1$ and $m<n\leq N$. Diagonalizing the positive matrix $a=(a_{l,m})$, Eq.~\eqref{eq2} can be rewritten as follows:
\begin{equation}\label{eq3}
\mathcal{L}_D(\rho)=\sum_k\gamma_k(L_k\rho L_k^\dagger-\frac{1}{2}\{L_k^\dagger L_k,\rho\}),
\end{equation}
where the parameters $\gamma_k$ are the eigenvalues of the matrix $(a_{l,m})$. After a Rotating Wave Approximation, the field-free Hamiltonian can be removed and we assume that the interaction Hamiltonian depends on $N_c$ time-dependent control fields, $u_k(t)$. The Hamiltonian $H_I$ can be expressed as $H_I=\sum_{k=1}^{N_c}u_k(t)H_k$, where $H_k$ are the different interaction terms. We make a standard controllability assumption for which any transformation of $SU(N)$ can be generated in an arbitrarily short time with respect to the relaxation times~\cite{schirmer01}. This hypothesis is verified if the Lie algebra generated by the Hermitian operators $H_k$ is $su(N)$ and if the maximum intensity of the control fields is very large with respect to the relaxation rates.

The quantum state $\rho$ can be expressed through a coherence vector $s$~\cite{alickibook,schirmer} whose coordinates $s_k$, $k=1,\cdots,N^2-1$ are the expectation values of the $N^2-1$ generalized Pauli matrices. The purity $p=\textrm{Tr}[\rho^2]$ of the density matrix is given by:
$$
p=\frac{1}{N}+\sum_{k=1}^{N^2-1}s_k^2=\frac{1}{N}+s^2,
$$
with $s^2=s\cdot s$. The map which sends $\rho$ to $s$ is an embedding from the space of density matrices to $\mathbb{R}^{N^2-1}$. The $N^2-N$ first components of the coherence vector can be written as the sum of off-diagonal terms of the density matrix, while the $N-1$ others depend on the diagonal elements (The case of a three-level quantum system is described in Appendix~\ref{secappb}). We denote by $s_o$ and $s_d$ the projections of $s$ on the two subspaces, the indices $o$ and $d$ being associated to off-diagonal and diagonal terms. We have $s=(s_o,s_d)$.

The Lindblad equation can be written in the coherence vector formalism as follows:
\begin{equation}\label{eqoct}
\begin{cases}
\dot{s}_o=R_o s_o+\sum_{k=1}^{N_c}u_k(t)(A_{oo}^{(k)}s_o+A_{od}^{(k)}s_d) \\
\dot{s}_d=q_d+R_ds_d+\sum_{k=1}^{N_c}u_k(t)(A_{do}^{(k)}s_o+A_{dd}^{(k)}s_d)
\end{cases}
\end{equation}
where the vector $q=(0,q_d)$ and the matrix $R$ represent respectively the inhomogeneous and homogeneous terms of the relaxation process. Note that $R$ is a block-diagonal matrix which does not coupled $s_o$ and $s_d$. $R_o$ is a diagonal matrix whose elements $\Gamma_{ij}$, $i\neq j$, are the dephasing rates of the transitions from level $i$ to $j$. The full matrix $R_d$ and the vector $q_d$ only depend on $\gamma_{ij}$, the rates of population relaxation from level $j$ to $i$~\cite{schirmer:04}. The block operator $A^{(k)}$ of components $(A_{oo}^{(k)},A_{od}^{(k)},A_{do}^{(k)},A_{dd}^{(k)})$ corresponds in this space to the interaction Hamiltonian $H_k$. Note that $A_{dd}$ is a zero matrix. The unitary dynamics are described by rotations on a sphere of radius $||s||$ and the generators $A^{(k)}$ are elements of the Lie algebra $so(N^2-1)$ of skew-symmetric matrices which verify $A^{(k)}=-{A^{(k)}}^\intercal$. However, all the rotations of $SO(N^2-1)$ cannot be realized by the set $\{A^{(k)}\}$ and only states belonging to the unitary orbit of the initial density matrix can be reached~\cite{schirmer}. At the density matrix level, this orbit is defined by the invariance of the spectrum of $\rho(t)$ by unitary dynamics. %Starting from a pure state, the orbit is $SU(N)/[SU(N-1)\times SU(1)]$ of real dimension $N^2-(N-1)^2-1=2(N-1)$, while the dimension of $SO(N^2-1)$ is $(N^2-1)(N^2-2)/2$. For $N=3$, the dimension of $SO(3^2-1)$ is 15 while the dimension of pure state orbit is only 4.

In order to be able to derive time-optimal trajectories, we introduce fictitious control fields so that any rotation of $SO(N^2-1)$ can be generated. This idea is the key point of the approach presented in this work. More precisely, instead of considering the optimal control problem defined by Eq.~\eqref{eqoct}, we now study the dynamical system controlled by $\tilde{N}_c>N_c$ fields such that $\textrm{Lie}[\{A^{(k)}\}_{k=1,\cdots, \tilde{N}_c}]=so(N^2-1)$. We deduce that any point of the hypersphere $||s||=s_f$ can be reached in an arbitrary small time from any other point. The increase in the number of controls available implies that the duration of the new process is less than the original control time and can be interpreted as a speed limit time of the problem.
\section{The general approach}\label{sec2new}
We show in this paragraph how to find the trajectories which optimize the rate of purity change of the quantum system. We have found more convenient to express the corresponding optimal control problem in a Lagrangian formalism.

We introduce a Lagrangian $\mathcal{L}$, which is defined as:
$$
\mathcal{L}=\frac{1}{2}\frac{d}{dt}s^2+\mu(s^2-s_f^2),
$$
where $\mu$ is a Lagrangian multiplier and $s_f$ a constant with $0\leq s_f^2\leq 1-\frac{1}{N}$. The Lagrangian $\mathcal{L}$ allows us to determine the coherence vector $s$ which optimizes the time evolution of the purity within the constraint of a fixed purity, $||s||=s_f$. The Lagrangian can be expressed as:
$$
\mathcal{L}=s^\intercal q +s^\intercal R s+\sum_k u_k s^\intercal A^{(k)} s+\mu (s^\intercal s-s_f^2).
$$
The maximization condition $\frac{\partial \mathcal{L}}{\partial s}=0$ leads to:
$$
q+(R+R^\intercal+2\mu) s=0,
$$
and does not depend on the control fields because $A^{(k)}$ is a skew-symmetric matrix. Decomposing the coordinates of the coherence vector, we arrive at:
\begin{equation}\label{eqcond}
\begin{cases}
(R_o+\mu)s_o=0 \\
q_d+(R_d+R_d^\intercal+2\mu) s_d=0
\end{cases}
\end{equation}
To simplify the discussion, we assume that all the dephasing rates are equal so that $R_o=-\Gamma I$, where $I$ is the identity matrix. If it is not the case then only the coordinates of $s_o$ associated to the maximum dephasing rate have to be accounted for. We deduce from Eq.~\eqref{eqcond} that $\mu=\Gamma$ or $s_o=0$. These two conditions define two geometric objects in the coherence vector space that are called magic subspaces.

The first one, $\mathcal{M}_d$, for which $s_o=0$ is a subspace of dimension $N-1$ and corresponds to diagonal density matrices. %In the case where $\textrm{det}[R_d+{R_d}^\intercal+2\mu I]\neq 0$, we deduce that $s_d=-(R_d+{R_d}^\intercal+2\mu I)^{-1}q_d$ and the Lagrange multiplier satisfies the relation:
%$$
%q_d^\intercal (R_d+{R_d}^\intercal+2\mu I)^{-2}q_d=s_f^2.
%$$
The second subspace $\mathcal{M}_o$ is characterized by the equation:
$$
q_d+(R_d+R_d^\intercal+2\Gamma I) s_d=0
$$
which gives, if $\textrm{det}[R_d+{R_d}^\intercal+2\Gamma I]\neq 0$, that
$$
s_d^{(m)}=-(R_d+{R_d}^\intercal+2\Gamma I)^{-1}q_d.
$$
This set is a subspace of dimension $N^2-N$ whose elements are density matrices with fixed diagonal coordinates. Note that this set is not empty only if $(s_d^{(m)})^2\leq 1-\frac{1}{N}$. In the limit $\Gamma\to +\infty$, we obtain $s_d^{(m)}\simeq -\frac{q_d}{2\Gamma}$. Since $q_d$ only depends on the relaxation rates $\gamma_{ij}$, it is straightforward to show that this subspace converges towards the set of density matrices with zero diagonal elements.

The next step consists in computing the time evolution of the system along the two magic subspaces. We introduce the relative purities $p_o=s_o^2$ and $p_d=s_d^2$, with $p_o+p_d=s^2$. On $\mathcal{M}_o$, we have $\dot{s}_d=0$ so the control fields depend only on $s_o$ and fulfill the following relation:
$$
\sum_ku_k^{(m)}(A_{do}^{(k)}s_o+A_{dd}^{(k)}s_d^{(m)})=-q_d-R_ds_d^{(m)},
$$
which leads to
\begin{equation}\label{eqmagic}
\sum_ku_k^{(m)}{s_d^{(m)}}^\intercal A_{do}^{(k)}s_o=-{s_d^{(m)}}^\intercal q_d-\frac{1}{2}{s_d^{(m)}}^\intercal (R_d+R_d^\intercal)s_d^{(m)}
\end{equation}
since $A_{dd}^{(k)}$ is a skew-symmetric matrix. Note that different trajectories can be followed on this space but the global evolution will not depend on this choice. Indeed, using Eq.~\eqref{eqoct}, it can be shown that:
$$
\dot{p}_o=2s_o^\intercal R_os_o-2\sum_k u_k^{(m)}{s_d^{(m)}}^\intercal A_{do}^{(k)}s_o
$$
which, from Eq.~\eqref{eqmagic}, transforms into:
$$
\dot{p}_o=-2\Gamma p_o+2{s_d^{(m)}}^\intercal q_d+{s_d^{(m)}}^\intercal (R_d+R_d^\intercal)s_d^{(m)}.
$$
It is worthwhile to mention here that all the coefficients of this differential equation can be expressed in terms of the relaxation parameters. The general solution can be written as:
$$
p_o(t)=p_o(0)e^{-2\Gamma t}+\frac{\lambda}{\Gamma}(1-e^{-2\Gamma t}),
$$
with $\lambda={s_d^{(m)}}^\intercal q_d+\frac{1}{2}{s_d^{(m)}}^\intercal (R_d+R_d^\intercal)s_d^{(m)}$. The purity $p_o$ is equal to zero when:
$$
t_o=\frac{1}{2\Gamma}\ln \big(\frac{\lambda-\Gamma p_o(0)}{\lambda}\big).
$$
Note that, since $p_o(t)$ decreases along the trajectory, we have $\lambda-\Gamma p_o(0)>0$. Here again, we can analyze the behavior of $t_o$ when $\Gamma\to +\infty$. In this limit, we have $\lambda\simeq -\frac{q_d^\intercal q_d}{2\Gamma}$. Starting from a pure state with $p_o(0)=1-\frac{1}{N}-(s_d^{(m)})^2$, we arrive at:
$$
t_o\simeq \frac{\ln \Gamma}{\Gamma}.
$$

The same analysis can be done on $\mathcal{M}_d$ where $s_o=0$. In this case, the goal is to determine the time evolution of the Lagrange multiplier $\mu(t)$. Along $\mathcal{M}_d$, we first have $s_d=-M^{-1}q_d$, where $M=R_d+R_d^\intercal+2\mu I$. Using $p_d=s_d^\intercal s_d$, we obtain:
$$
\dot{p}_d=2s_d^\intercal \dot{s}_d=q_d^\intercal-2\mu s_d^\intercal s_d.
$$
However, the time derivative of $p_d$ can also be expressed as:
$$
\dot{p}_d=\dot{\mu} \frac{d}{d\mu} p_d
$$
where $\frac{d}{d\mu}$ denotes the derivative with respect to $\mu$. Since $\frac{d}{d\mu} s_d=-2 M^{-1}s_d$, we finally get:
\begin{equation}\label{eqmutgen}
\dot{\mu}=\frac{2\mu s_d^\intercal s_d-q_d^\intercal s_d}{4s_d^\intercal M^{-1}s_d}.
\end{equation}
Integrating analytically or numerically Eq.~\eqref{eqmutgen}, we obtain the time evolution of $\mu$ in $\mathcal{M}_d$, and therefore the evolution of $s_d$ and $p_d$ in this space. This approach will be used in Sec.~\ref{sec3} and \ref{sec4} for two- and three- level quantum systems.
\section{The case of a two-level quantum system}\label{sec3}
We analyze in this paragraph the evolution of the purity in a dissipative two-level quantum system. Since no control parameter is added in this case, the general approach developed in Sec.~\ref{sec2new} allows us to recover the results established in~\cite{lapert,lapert:2013} by optimal control theory. The lower bound for a two-level quantum system corresponds exactly to the minimum time of the control process and is therefore tight. %A comparison is also made between this minimum time and the two bounds proposed in~\cite{kosloff} whose definition is recalled in Appendix~\ref{secappa}.

In the Bloch representation, the equations of motion of the coherence vector $s=(s_1,s_2,s_3)$ can be expressed as:
$$
\begin{cases}
\dot{s}_1= - \Gamma s_1+u_2 s_3 \\
\dot{s}_2= -\Gamma s_2-u_1 s_3 \\
\dot{s}_3=\gamma_{-}-\gamma_{+}s_3+u_1s_2 - u_2s_1
\end{cases}
$$
where $\gamma_-=\gamma_{12}-\gamma_{21}$ and $\gamma_+=\gamma_{12}+\gamma_{21}$. The dephasing rate $\Gamma$ fulfills the constraint $\Gamma\geq \frac{\gamma_+}{2}$~\cite{schirmer:04}. The system is controlled by two time-dependent fields, $u_1$ and $u_2$. The coordinates of the equilibrium point of the dynamics are $(0,0,s_3^{(e)}=\frac{\gamma_-}{\gamma_+})$. To simplify the description of the solution, we assume below that $\gamma_->0$, i.e. $s_3^{(e)}>0$.

\begin{figure}[htp]
\centering
\includegraphics[width=1.0\textwidth]{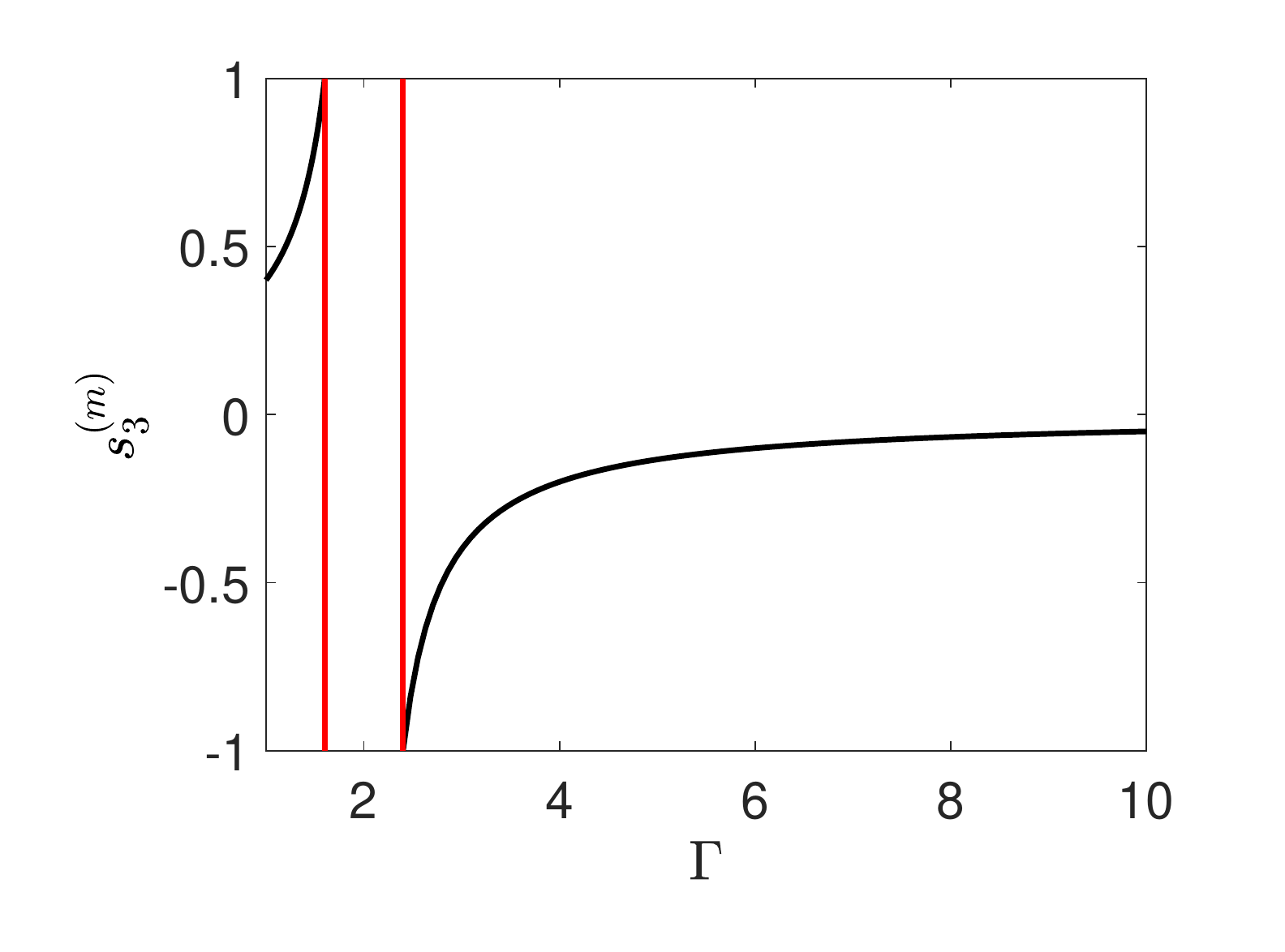}
\caption{(Color online) Position of the magic plane $s_3^{(m)}$ as a function of $\Gamma$. The parameters are set to $\gamma_+=2$ and $\gamma_-=0.8$, with $\Gamma\geq\frac{\gamma_+}{2}$. The two vertical red (or dark gray) lines delimit the values of $\Gamma$ for which there is no intersection between the magic plane and the Bloch ball.}
\label{fig3}
\end{figure}

We first apply the general theory to find the magic subspaces. The coordinates of the coherence vector $s$ can be decomposed into $s_o=(s_1,s_2)$ and $s_d=(s_3)$. The Lagrangian $\mathcal{L}$ can be expressed as:
$$
\mathcal{L}=-\Gamma s_1^2-\Gamma s_2^2+\gamma_-s_3-\gamma_+s_3^2+\mu (s_1^2+s_2^2+s_3^2-s_f^2).
$$
The extremal conditions are given by:
$$
\begin{cases}
(\Gamma-\mu)s_1=0 \\
(\Gamma-\mu)s_2=0 \\
\gamma_--2\gamma_+s_3+2\mu s_3=0
\end{cases}
$$
We deduce that there are two magic subspaces. The first one $\mathcal{M}_o$, a plane for which $\mu=\Gamma$, is characterized by a fixed value of $s_3=s_3^{(m)}$:
$$
s_3^{(m)}=\frac{-\gamma_-}{2(\Gamma-\gamma_+)}.
$$
Using the constraint $\Gamma\geq \frac{\gamma_+}{2}$, we deduce that $s_3^{(m)}\in [-1,0[$ for $\Gamma\in [\gamma_++\frac{\gamma_-}{2},+\infty[$ and $s_3^{(m)}\in [s_3^{(e)},1]$ if $\Gamma\in [\frac{\gamma_+}{2},\gamma_+-\frac{\gamma_-}{2}]$. The position of the different magic planes as a function of $\Gamma$ is displayed in Fig.~\ref{fig3}.
A trajectory lies on this plane if $u_1$ and $u_2$ satisfy:
$$
\gamma_{-}-\gamma_{+}s_3^{(m)}+u_1s_2 - u_2s_1=0.
$$
A solution is for instance given by $u_2=\frac{\gamma_{-}-\gamma_{+}s_3^{(m)}}{s_1}$~\cite{lapert,lapert:2013}.

The second magic space, $\mathcal{M}_d$, corresponds to the $s_3$- axis, with $s_1=s_2=0$. In this case, we have $s_3=\frac{-\gamma_-}{2(\mu-\gamma_+)}$ and the Lagrange multiplier is determined by the condition $s_3^2=\frac{\gamma_-^2}{4(\mu-\gamma_+)^2}=s_f^2$. Any point of the $s_3$- axis can be reached when $\mu \in ]-\infty,\gamma_+-\frac{\gamma_-}{2}]\cup [\gamma_++\frac{\gamma_-}{2},+\infty[$. We can move along this space with zero control fields.

\begin{figure}[htp]
\centering
\includegraphics[width=1.0\textwidth]{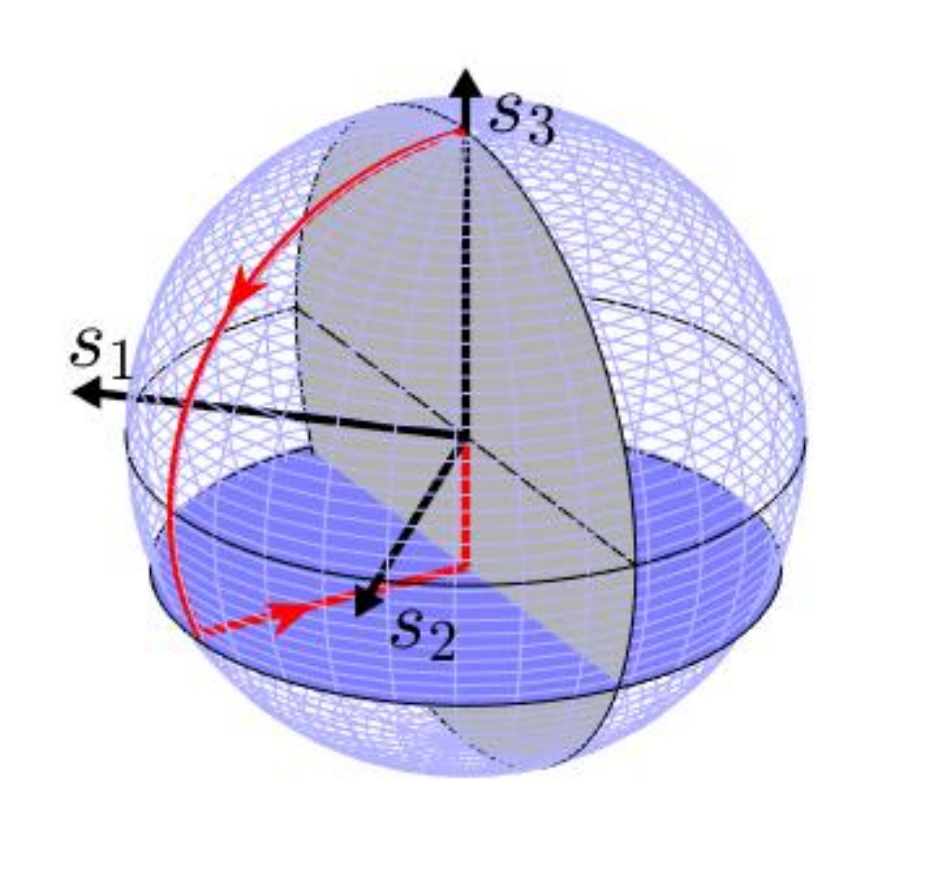}
\caption{(Color online) Time-optimal trajectory (red or light gray solid line) to reach the center of the Bloch ball starting from the north pole (equilibrium point of the dynamics). The blue (dark gray) horizontal plane is the magic plane of equation $s_3=s_3^{(m)}$, which is parallel to the equatorial plane. The initial state is the north pole of the Bloch sphere. We consider the case where $\gamma_-=\gamma_+$.}
\label{fig1}
\end{figure}

As an illustrative example, we consider a control process which is aimed at steering the system from the equilibrium state to the center of the Bloch ball of coordinates $(0,0,0)$, i.e. the completely mixed state. This control process can find applications in Nuclear Magnetic Resonance~\cite{lapert} or in quantum computing. The goal is therefore to decrease the purity of the system as fast as possible. Note that the same analysis could be done for any other points of the Bloch ball. The time evolution of the purity on the two magic subspaces can be written as
$$
\dot{p}_o=-2\Gamma p_o+2\gamma_- s_3^{(m)}-2\gamma_+ (s_3^{(m)})^2
$$
for $\mathcal{M}_o$ and
$$
\dot{p}_d=2\gamma_-s_3-2\gamma_+s_3^2,
$$
for $\mathcal{M}_d$. It can be shown that the fastest way to shrink the purity is to follow a path along $\mathcal{M}_o$~\cite{lapert}. We therefore deduce that the optimal trajectory is the concatenation of an arc of circle along the Bloch sphere to reach the magic plane, followed by a path onto this space up to the $s_3$- axis where $p_o=0$ and an arc along this axis. Since there is no limitation on the maximum intensity of the control fields, the initial time to reach the magic plane is negligible. A time-optimal trajectory is represented in Fig.~\ref{fig1}.
The last step of the method consists in computing the corresponding control time. Along $\mathcal{M}_o$, the purity evolves as:
$$
p_o(t)=p_o(0)e^{-2\Gamma t}+2(\gamma_--\gamma_+s_3^{(m)})\frac{s_3^{(m)}}{\Gamma}(1-e^{-2\Gamma t}),
$$
with $p_o(0)=\frac{\gamma_-^2}{\gamma_+^2}-\frac{\gamma_-^2}{4(\Gamma-\gamma_+)^2}$. We then deduce the time $t_o$:
$$
t_o=\frac{1}{2\Gamma}\ln \big(1+\frac{2p_o(0)\Gamma(\Gamma-\gamma_+)^2}{\gamma_-^2(2\Gamma-\gamma_+)}\big).
$$
There are two different ways to derive the time $t_d$ to go along the $s_3$- axis from $s_3^{(m)}$ to 0. The simplest approach consists in using the fact that
the two control fields are zero. Since $\dot{s}_3=\gamma_--\gamma_+s_3$, we deduce that:
$$
t_d=\frac{1}{\gamma_+}\ln\big(\frac{2\Gamma-\gamma_+}{2(\Gamma-\gamma_+)}\big).
$$
The second method is based on the computation of the time evolution of $\mu$ as explained in Sec.~\ref{sec2new}. This approach is described in Appendix~\ref{secappc}.

The total minimum time $t_{MS}$ is finally given by $t_{MS}=t_o+t_d$. In the limit $\Gamma\gg \gamma_+$, this time can be approximated as:
$$
t_{MS}\simeq_{\Gamma\gg\gamma_+} \frac{\ln \Gamma}{\Gamma}
$$
\section{Application to a three-level quantum system}\label{sec4}
We consider in this paragraph the example of a three-level quantum system and the same control problem as in Sec.~\ref{sec3}. We denote by 1, 2 and 3 the three energy levels. We assume that the non-zero relaxation rates are given by:
$$
\gamma_{12}=1,~\gamma_{13}=0.5,~\gamma_{23}=0.5.
$$
The coherence rates satisfy $\Gamma_{ij}=\tilde{\Gamma}_{ij}+\frac{\gamma_{ij}+\gamma_{ji}}{2}$ where $\tilde{\Gamma}_{ij}$ denote the pure dephasing terms which fulfill the inequalities~\cite{schirmer:04}:
$$
(\sqrt{\tilde\Gamma_b}-\sqrt{\tilde\Gamma_a})^2\leq \tilde\Gamma_a \leq (\sqrt{\tilde\Gamma_b}+\sqrt{\tilde\Gamma_c})^2,
$$
where the indices $a$, $b$ and $c$ are any permutation of $12$, $13$ and $23$. We choose the parameter $\tilde{\Gamma}_{ij}$ so that $\Gamma_{ij}$ is the same for all the energy-level transitions. An explicit derivation of the coherence vector dynamics is given in Appendix~\ref{secappb}. In a compact form, we obtain:
$$
\begin{cases}
\dot{s}_o = -\Gamma I+\sum_k u_k(A_{oo}s_o+A_{od}s_d) \\
\dot{s}_d = q_d+R_d s_d+\sum_k u_k(A_{do}s_o+A_{dd}s_d),
\end{cases}
$$
where $s_o$ and $s_d$ are respectively a six and a two dimensional vectors of coordinates $(s_1,s_2,\cdots,s_6)$ and $(s_7,s_8)$. We denote by $(q_7,q_8)$ the components of $q_d$ and by:
$$
R_d=\begin{pmatrix}
r_{77} & r_{78} \\
r_{87} & r_{88}
\end{pmatrix}
$$
the ones of $R_d$ which can be expressed as a function of the relaxation rates $\gamma_{ij}$. We now follow the general procedure presented in Sec.~\ref{sec2new} and we introduce the Lagrangian $\mathcal{L}$:
\begin{eqnarray*}
& &\mathcal{L}=-\Gamma \sum_{k=1}^6 s_k^2+q_7s_7+q_8s_8+s_7r_{77}s_7+s_8r_{88}s_8\\
& &+s_7(r_{78}+r_{87})s_8+\mu (s^2-s_f^2).
\end{eqnarray*}
The magic subspaces are the subspace of diagonal density matrices such that $s_o=0$ and the subspace defined by $\frac{\partial\mathcal{L}}{\partial s_7}=0=\frac{\partial\mathcal{L}}{\partial s_8}$. This leads to:
\begin{equation}\label{eqmagic3}
\begin{cases}
q_7+2r_{77}s_7+(r_{78}+r_{87})s_8+2\Gamma s_7 =0 \\
q_8+2r_{88}s_8+(r_{78}+r_{87})s_7+2\Gamma s_8=0
\end{cases}
\end{equation}
with $\mu=\Gamma$. Equation~\eqref{eqmagic3} gives the position of the six-dimensional magic subspace defined by $s_7^{(m)}$ and $s_8^{(m)}$. For $\Gamma=2$, we deduce that:
$$
s_7^{(m)}=-0.1928;~s_8^{(m)}=-0.1485,
$$
which leads to:
$$
\rho_{11}=0.1364;~\rho_{22}=0.4091;~\rho_{33}=0.4545.
$$
Starting from a purity equal to one at time $t=0$, the time spent along this space such that $p_o(t_o)=0$ is:
$$
t_o=\frac{1}{2\Gamma}\ln \big(\frac{\lambda-\Gamma p_o(0)}{\lambda}\big)\simeq 0.5613.
$$

We now determine the time to go from $\mathcal{M}_o$ to the zero coherence vector. We follow the general approach. The details can be found in Appendix~\ref{secappc}. It can be shown that the Lagrange multiplier fulfills the following equation:
\begin{equation}\label{eqmut3}
\dot{\mu}=\frac{q_7s_7+q_8s_8-2\mu (s_7^2+s_8^2)}{2(s_7\frac{d s_7}{d\mu}+s_8\frac{d s_8}{d\mu})},
\end{equation}
where $s_7$ and $s_8$ are two functions of $\mu$ as displayed in Fig.~\ref{fig4a}. The explicit expression is given in Eq.~\eqref{eqs78} of Appendix~\ref{secappc}.
\begin{figure}[htp]
\centering
\includegraphics[width=1.0\textwidth]{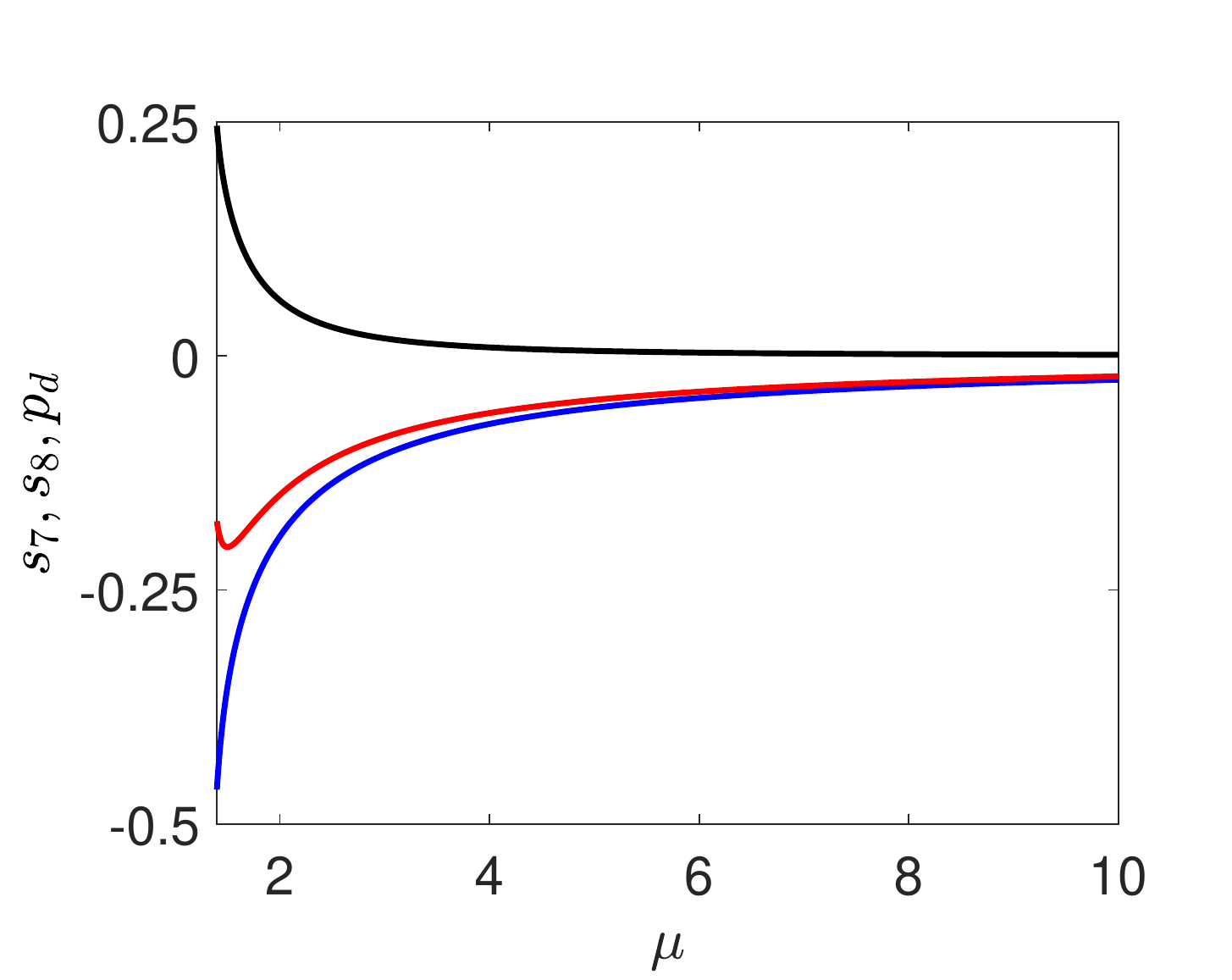}
\caption{(Color online) Evolution of $p_d$ (black), $s_7$ (blue or dark gray) and $s_7$ (red or light gray) as a function of $\mu$. The parameter $\mu$ belongs to the interval $[1.4,10]$. The parameter $\Gamma$ is set to 2.}
\label{fig4a}
\end{figure}
Equation~\eqref{eqmut3} can be integrated numerically. The time evolution of $\mu$ is represented in Fig.~\ref{fig4b} in the case $\Gamma=2$. By construction, the initial value of $\mu$ is $\Gamma$. We observe that $\mu$ diverges for a finite time of the order of 0.337. The coherence vector is zero at this time.
\begin{figure}[htp]
\centering
\includegraphics[width=1.0\textwidth]{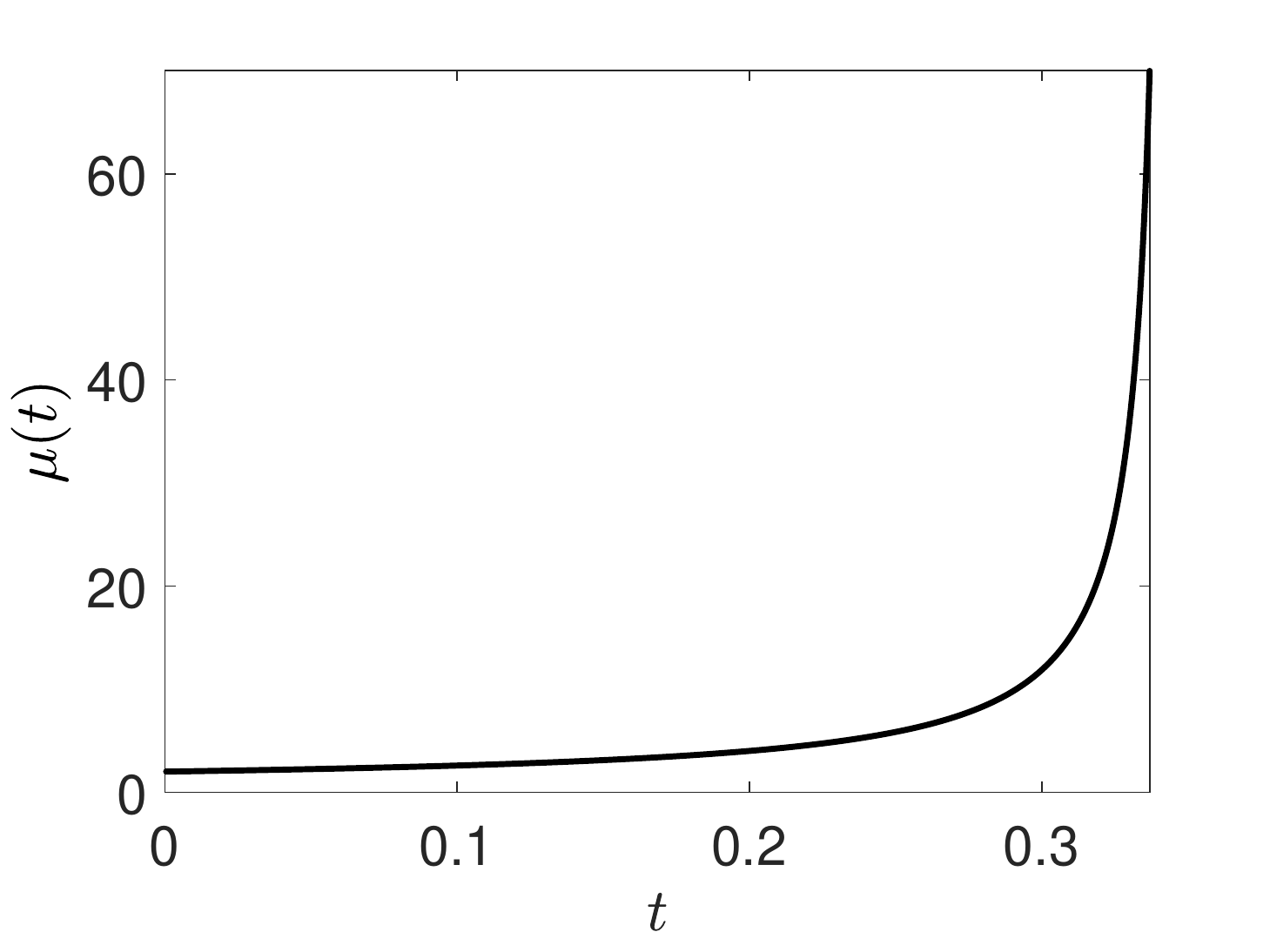}
\caption{(Color online) Time evolution of $\mu$ in the interval $[0,0.337]$ for $\Gamma=2$. The value of $\mu$ at time 0 is taken to be $\Gamma$.}
\label{fig4b}
\end{figure}
We finally plot in Fig.~\ref{fig5} the evolution of the minimum time $t_{MS}$ predicted by the magic subspace approach as a function of $\Gamma$. We show that $t_{MS}$ can be well approximated by $\ln\Gamma/\Gamma$ when $\Gamma\geq 10$. Since this approximation is less than $t_{MS}$, it can be used as a lower bound to the original minimum time of the control process.

\begin{figure}[htp]
\centering
\includegraphics[width=1.0\textwidth]{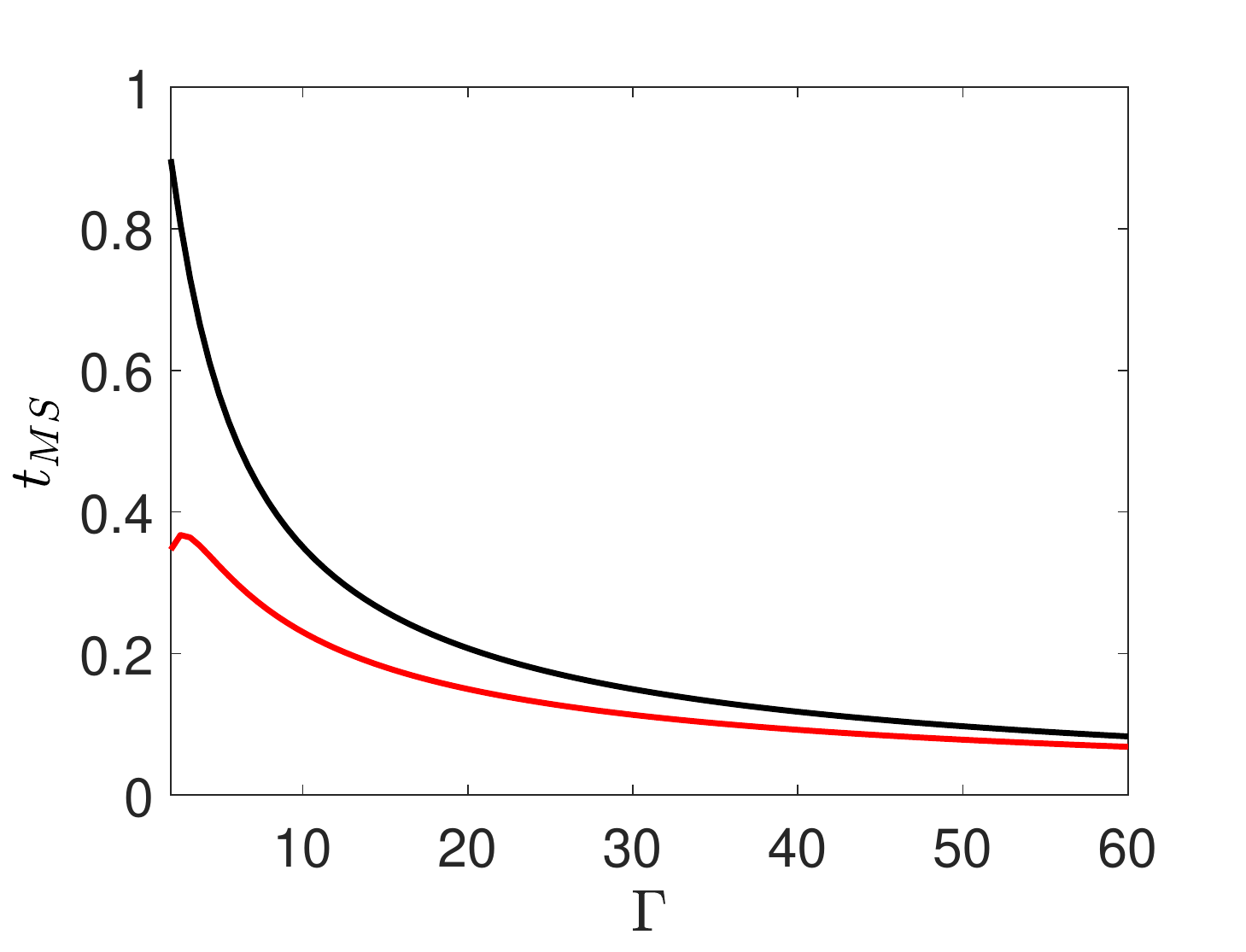}
\caption{(Color online) Evolution of the minimum time $t_{MS}$ as a function of $\Gamma$ (black line). The red (dark gray) line depicts the time $\frac{\ln \Gamma}{\Gamma}$, which is a good approximation of the minimum time when $\Gamma \gg 1$.}
\label{fig5}
\end{figure}

\section{Comparison of Purity Speed Limits}\label{sec5}
This section is aimed at comparing the speed limit derived in this study with the ones of Ref.~\cite{kosloff}. The minimum time is also estimated by using a numerical optimal control algorithm~\cite{grape}.

\begin{figure}[htp]
\centering
\includegraphics[width=1.0\textwidth]{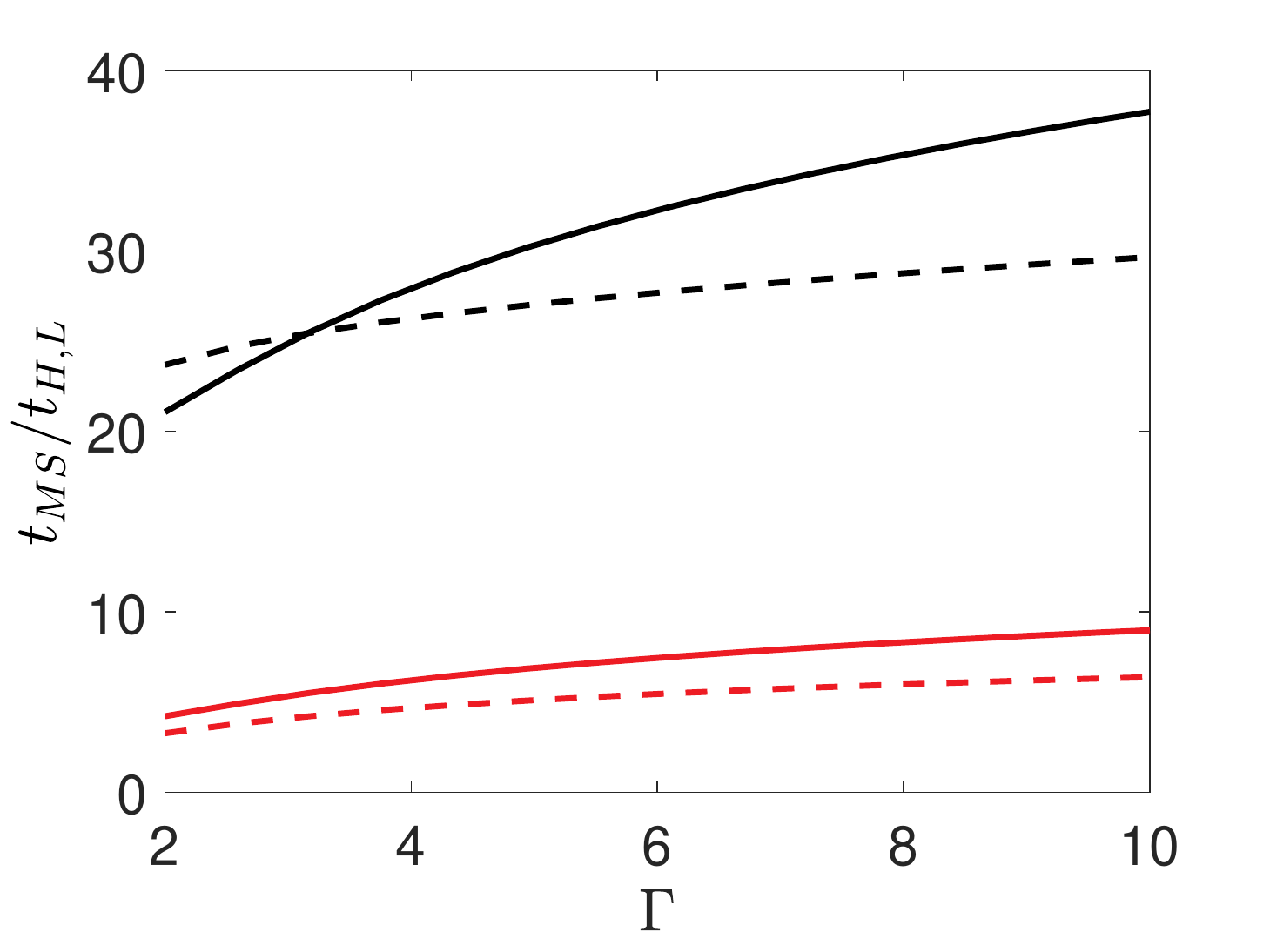}
\caption{(Color online) Evolution of the ratios $t_{MS}/t_H$ (black line) and $t_{MS}/t_L$ (red or light gray line) as a function of $\Gamma$ for two- (solid line) and three- (dashed line) level quantum systems. Numerical parameters for the two-level quantum system are set to $\gamma_+=1$, $\gamma_-=0.5$ and $\Gamma\geq 2$.}
\label{fig6}
\end{figure}

Two PSL have been established in~\cite{kosloff} based on a decomposition of the Lindblad operator either in the Hilbert or in the Liouville space. The definition and the derivation of the two PSL are recalled in Appendix~\ref{secappa}. We denote by $t_H$ and $t_L$, the two bounds on the minimum control time. For a two-level quantum system, we get:
$$
\begin{cases}
t_H=\frac{\ln 2}{4[\Gamma+\frac{\gamma_+}{2}+|\gamma_-|]} \\
t_L=\frac{\ln 2}{\max(2\Gamma,\gamma_++\sqrt{\gamma_+^2+\gamma_-^2})}.
\end{cases}
$$
while for the three- level system analyzed in Sec.~\ref{sec4}, we have:
$$
\begin{cases}
t_H=\frac{\ln 3}{16+\frac{4\sqrt{3}}{3}+4|\Gamma-\frac{5}{6}|+4|\Gamma-\frac{1}{2}|} \\
t_L=\frac{\ln 3}{\max(2\Gamma,1+\frac{\sqrt{10}}{2})}.
\end{cases}
$$
Note that, for $t_H$, we use here the basis of the normalized Pauli matrices. The tightness of a speed limit represents how precisely the corresponding time bounds the actual minimum time spent by the system to reach a suitable target state. A measure of the tightness is given by $t_{MS}/t_{L, H}$ for a two-level quantum system. We consider also this ratio for higher-dimensional systems to estimate the gain obtained from the speed limit of this study. Figure~\ref{fig6} displays the evolution of this measure as a function of $\Gamma$. As expected, $t_L$ is a better bound than $t_H$, but a large ratio is observed for the two PSL. Such results show on these two examples the interest of the speed limit formulation presented in this work. The same conclusion holds true in the general case of a $N$- level quantum system when $\Gamma\gg 1$. Indeed, a rapid analysis of $t_H$ and $t_L$ shows that they evolve, up to a constant factor, as $\frac{1}{\Gamma}$ in this limit, while $t_{MS}$ is of the order of $\frac{\ln(\Gamma)}{\Gamma}$. More precisely, for a $N$- level quantum system, we have:
$$
\begin{cases}
t_H\simeq_{\Gamma\gg 1} \frac{\ln(N)}{2^N\Gamma} \\
t_L\simeq_{\Gamma\gg 1} \frac{\ln(N)}{2\Gamma}
\end{cases}
$$
while $t_{MS}\simeq_{\Gamma\gg 1}\frac{\ln(\Gamma)}{\Gamma}$. For a fixed number of levels, the corresponding ratio, which goes as $\ln(\Gamma)$, diverges. Note also that the limit of $t_{MS}$ does not depend on the number of levels $N$.

\begin{figure}[htp]
\centering
\includegraphics[width=1.0\textwidth]{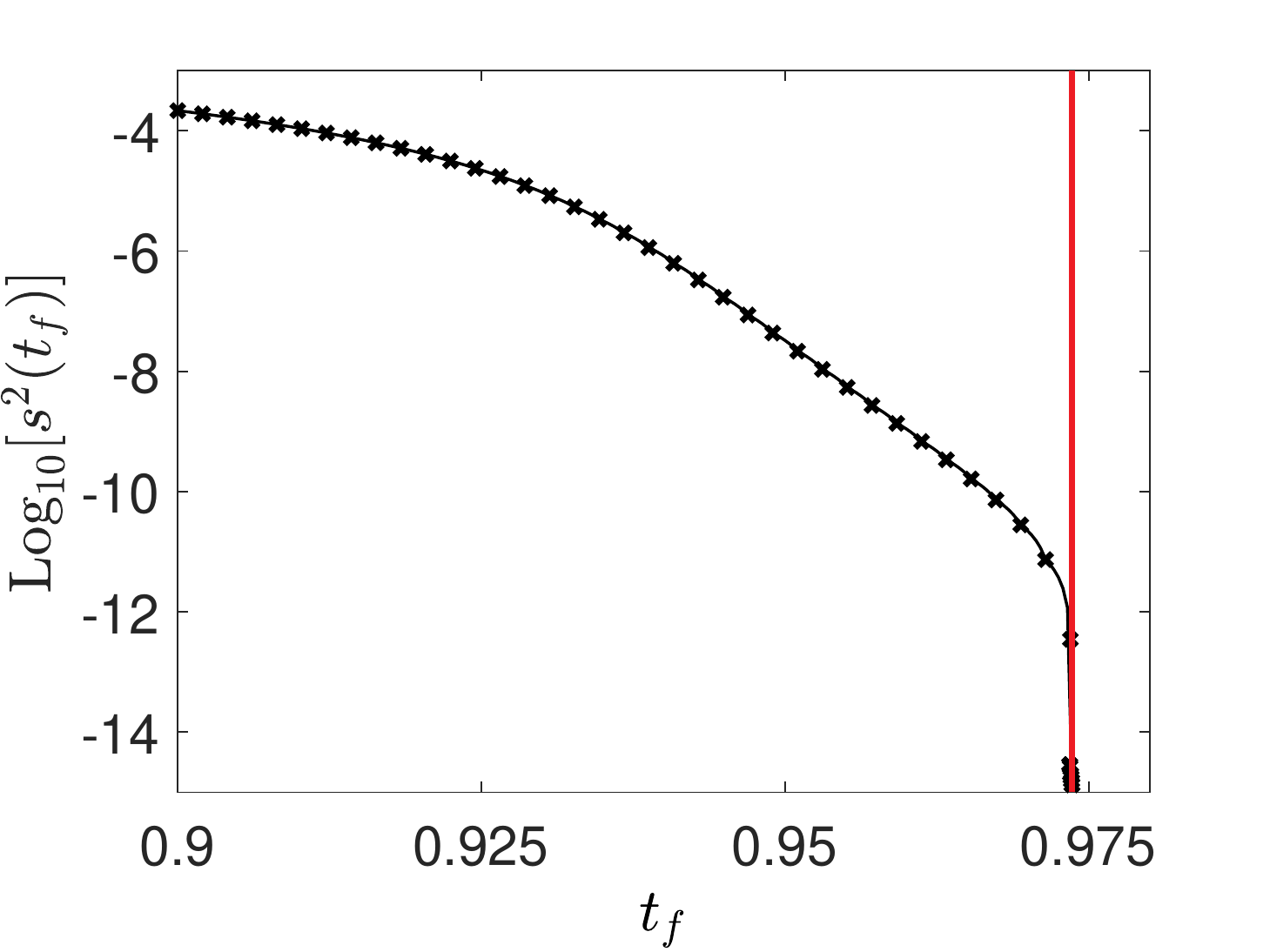}
\caption{(Color online) Evolution of the square modulus of the coherence vector (crosses) generated by numerical optimization as a function of the control time $t_f$ in the case of the three-level quantum system. The solid black line is just to guide the eye. The vertical line in red (or dark gray) indicates the minimum time, which is estimated to be of the order of 0.9735.}
\label{fig7}
\end{figure}

In the case of the three-level quantum system with $\Gamma=2$, we finally present numerical optimization results in order to estimate the minimum control time $t^*$ in the original control problem. We consider a gradient algorithm, GRAPE, which has been described in detail elsewhere~\cite{grape}. We start from a point of $\mathcal{M}_o$ with a purity equal to 1. The goal is to reach the zero coherence vector in a fixed control time $t_f$. The cost functional to minimize is $s^2(t_f)$, i.e. the final square modulus of the coherence vector. There is no bound on the control fields. The computations are done for different control durations. As can be seen in Fig.~\ref{fig7}, we observe that the value of the cost function decreases as $t_f$ increases. At a certain control time, the pulse performance is numerically saturated. The corresponding time $t_f$ can be regarded as the minimum time $t^*$ of the control process. This time is estimated to be of the order of 0.9735. For the same control problem, the different speed limit times are $t_{MS}\simeq 0.8985$, $t_L\simeq 0.275$ and $t_H\simeq 0.038$. We observe that $t_{MS}$ gives a much better estimation of the minimum time, with an error of the order of 8\%.

\section{Conclusion}\label{sec6}
In this study, we have introduced a new approach for finding purity speed limits in dissipative quantum systems. The basic idea consists in enlarging the number of control available in order to connect two density matrices with the same purity. In a standard unitary framework, only a density matrix with the same spectrum as the initial state can be
reached. Such fictitious fields have the key advantage to simplify the corresponding time-optimal control problem. If there is no constraint on the maximum intensity of the fields, we show that the time-optimal trajectories belong to two magic subspaces, which can be defined in the coherence vector formalism. The two- and three- level cases have been discussed. The bound derived in this study is tight for two-level quantum systems because it corresponds exactly to the time obtained by optimal control theory. For a specific three-level quantum system, we have estimated that the error with respect to the minimum time is of the order of few percents. This work can therefore be viewed as a step forward in the understanding of the link between QSL and optimal control. It also opens the way to studies in the same direction in which the number of control fields is enlarged to determine the minimum time to control a given process. Finally, we have also shown the superiority of this bound with respect to other speed limits published in the literature. Finally, it would be interesting to explore the potential applications of this study in quantum thermodynamics or quantum computing in which the concept of QSL plays a key role.\\

\noindent\textbf{ACKNOWLEDGMENT}\\
D. Sugny acknowledges support from the QUACO project (ANR 17-CE40-0007-01). This project has received funding from the European Union's Horizon 2020 research and innovation programme under the Marie-Sklodowska-Curie grant agreement No 765267 (QUSCO).

\appendix

\section{Derivation of Purity Speed Limits}\label{secappa}
We recall in this paragraph the definition of the two PSL derived in~\cite{kosloff}. We consider a $N$- level quantum system whose dynamics are governed by the Lindblad equation~\eqref{eq1}.

Using the Frobenius norm of an operator $A$ defined by: $||A||=\sqrt{\textrm{Tr}(AA^\dagger)}$, a first purity speed limit in Hilbert space can be derived. We denote by $t_H$, a lower bound on the minimum time evolution. We have:
$$
t_H=\frac{|\ln(p(t_f)/p(0))|}{4\sum_{k,k'}^{N^2-1} |a_{k,k'}|\times ||V_k||\times ||V_{k'}||},
$$
where $p(0)$ and $p(t_f)$ are the initial and final purities of the system. Note that this bound depends on the operator basis used to express the Lindblad generator. This point is clarified below for the case of a two-level quantum system.

The Lindblad equation~\eqref{eq1} can be written in a Schr\"odinger-like form:
$$
i\frac{\partial}{\partial t}|\rho\rangle =\mathcal{H}|\rho\rangle,
$$
where the density matrix $\rho$ is written as a column vector and denoted $|\rho\rangle$, and $\mathcal{H}$ is the Hamiltonien superoperator of the dynamics. A second PSL can be established in this Liouville formalism and leads to the bound $t_L$, which can be expressed as:
$$
t_L=\frac{|\ln(p(t_f)/p(0))|}{||\mathcal{H}-\mathcal{H}^\dagger||_{SP}},
$$
where $SP$ means the spectral norm, i.e. the largest absolute value of the eigenvalues of the operator. Note that $t_L\geq t_H$, so the Liouville speed limit is always tighter than the Hilbert one.

We now derive the expression of the two speed limits in the case of two and three-level quantum systems. The computation can be done in the same way for higher-dimensional spaces.

For two-level systems, we consider the same notations as in Sec.~\ref{sec3}. In the basis of the normalized Pauli matrices, the matrix $a$ is given by:
\begin{equation}\label{hmatrixpauli}
a= \begin{pmatrix}
2\Gamma -\gamma_{+} & 0                         &  0                                                                      \\
0                                                                    &  \gamma_{+} &  -\dfrac{i\gamma_{-}}{2} \\
0                                                                    & \dfrac{i\gamma_{-}}{2} & \dfrac{\gamma_{+}}{2}
\end{pmatrix}
\end{equation}
which leads to:
\begin{equation}\label{eq::QSLHilbert2LevelPauli}
t_H= \frac{|\ln(p(t_f)/p(0))|}{4\left[ |\gamma_{-}| + \tfrac{\gamma_{+}}{2}+\Gamma\right] }
\end{equation}
The diagonal form of the Lindblad operator given by Eq.~\eqref{eq3} is defined by:
$$
L_1=\begin{pmatrix}
0 & 1 \\
0 & 0
\end{pmatrix};
~L_2=\begin{pmatrix}
0 & 0 \\
1 & 0
\end{pmatrix};~
L_3=\frac{1}{\sqrt{2}}\begin{pmatrix}
1 & 0 \\
0 & -1
\end{pmatrix}
$$
with $\gamma_1=\gamma_{21}$, $\gamma_2=\gamma_{12}$ and $\gamma_3=\Gamma-\frac{\gamma_+}{2}$. We therefore deduce that the bound can be expressed as:
\begin{equation}
t_H = \frac{|\ln(p(t_f)/p(0))|}{4\Gamma+\frac{\gamma_+}{2}},
\end{equation}
which shows on this example that the bound depends on the basis used to express the Lindblad operator.

In the Liouville space formalism , the dissipative part of the Hamiltonian $\mathcal{H}$ is:
$$
 \begin{pmatrix}
    -i \gamma_{21} & 0             & 0                 &  i\gamma_{12} \\
    0         & - i\Gamma  & 0                                  &  0       \\
    0  & 0                                    & -i \Gamma  & 0  \\
    i\gamma_{21}   & 0             & 0                 & -i\gamma_{12}
   \end{pmatrix}
$$
whose spectral norm is equal to:
$$
||\mathcal{H}||_{SP}=\max\left(2\Gamma, \gamma_++\sqrt{\gamma_+^2+\gamma_-^2} \right)
$$
and we deduce the corresponding lower bound:
\begin{equation}\label{eq::QSLLiouville2Level}
t_L= \frac{|\ln(p(t_f)/p(0))|}{\max\left(2\Gamma, \gamma_++\sqrt{\gamma_+^2+\gamma_-^2} \right)}
\end{equation}
%In the case of a three-level quantum system, we have to find the eigenvalues of the matrix $\gamma$:
%$$
%\gamma=\begin{pmatrix}
%-2(\gamma_{21}+\gamma_{31}) & \gamma_{12}+\gamma_{21} & \gamma_{13}+\gamma_{31} \\
%\gamma_{12}+\gamma_{21} & -2(\gamma_{12}+\gamma_{32}) & \gamma_{23}+\gamma_{32} \\
%\gamma_{13}+\gamma_{31} & \gamma_{23}+\gamma_{32} & -2(\gamma_{13}+\gamma_{23})
%\end{pmatrix}
%$$
%and the minimum time is:
%\begin{equation}\label{eq::QSLLiouville2Level}
%t_L= \frac{|\ln(p(t_f)/p(0))|}{\max\left(2\{\Gamma_{ij}\}_{i,j}, ||\gamma||_{SP} \right)},
%\end{equation}
%where $\{\Gamma_{ij}\}_{i,j}$ is the set of the parameters $\Gamma_{ij}$ with $i\neq j$.

In the case of a three-level quantum system, we have considered the following $a$- matrix with the shorthand notation:
$a_{-}=\gamma_{12}-\gamma_{21}$, $a_{+}=\gamma_{12}+\gamma_{21}$, $b_{+}=\gamma_{13}+\gamma_{31}$, $b_{-}=\gamma_{13}-\gamma_{31}$ , $c_{-}=\gamma_{23}-\gamma_{32}$, $c_{+}=\gamma_{23}+\gamma_{32}$ and $ X=\tfrac{\sqrt{3}}{6}(a_{-}-\gamma_{31}+\gamma_{32})$. We have:
%to find the eigenvalues of the matrix $\gamma$:
%$$
%\gamma=\begin{pmatrix}
%-2(\gamma_{21}+\gamma_{31}) & \gamma_{12}+\gamma_{21} & \gamma_{13}+\gamma_{31} \\
%\gamma_{12}+\gamma_{21} & -2(\gamma_{12}+\gamma_{32}) & \gamma_{23}+\gamma_{32} \\
%\gamma_{13}+\gamma_{31} & \gamma_{23}+\gamma_{32} & -2(\gamma_{13}+\gamma_{23})
%\end{pmatrix}
%$$
%and the minimum time is:
%\begin{equation}\label{eq::QSLLiouville2Level}
%t_L= \frac{|\ln(p(t_f)/p(0))|}{\max\left(2\{\Gamma_{ij}\}_{i,j}, ||\gamma||_{SP} \right)},
%\end{equation}
%where $\{\Gamma_{ij}\}_{i,j}$ is the set of the parameters $\Gamma_{ij}$ with $i\neq j$.

\begin{equation}
a=
\begin{pmatrix}
\mathbf{M}_{3\times 3}   & \mathbf{0}_{3\times 2}                        &  X \mathbf{e}_{3}(1,1)\\
\mathbf{0}_{2\times 3}    & \tfrac{1}{2}b_{+}\mathbf{1}_2 + \tfrac{\sqrt{2}}{2}b_{-}\mathbf{\sigma}_2  &  \mathbf{0}_{2\times 3} \\
X \mathbf{e}_{3}(1,1)  &  \mathbf{0}_{3\times 2} & \mathbf{M}'_{3\times 3}
\end{pmatrix}
\end{equation}
The matrix $\mathbf{M}_{3\times 3} $ is given by Eq.~\eqref{eq:AppA::M33} where $W=\tfrac{1}{2}(a_{+}+\gamma_{32}+\gamma_{31})$:
\begin{equation}\label{eq:AppA::M33}
\mathbf{M}_{3\times 3}=
\begin{pmatrix}
\Gamma-W &           0                     &            0                   \\
0                                                                                               & \tfrac{1}{2}a_{+}   &  -\tfrac{i}{2}a_{-} \\
0                                                                                               &\tfrac{i}{2}a_{-}     &  \tfrac{1}{2}a_{+}
\end{pmatrix}
\end{equation}
Also, $\mathbf{e}_{N}(i,j)$ represents a $N\times N$ matrix with $1$ in the $(i,j)$- entry and $0$ elsewhere. The matrix  $\mathbf{M}'_{3\times 3} $ is given by Eq.~\eqref{eq:AppA::M33prime} with $Y=\tfrac{1}{6}(a_{+}+\gamma_{31}+\gamma_{32})+\tfrac{2}{3}(\gamma_{13}+\gamma_{23})$:
\begin{equation}\label{eq:AppA::M33prime}
\mathbf{M}'_{3\times 3}=
\begin{pmatrix}
\Gamma-Y&           0                     &            0                   \\
0               & \tfrac{1}{2}c_{+}   &  -\tfrac{i}{2}c_{-} \\
0               &\tfrac{i}{2}c_{-}     &  \tfrac{1}{2}c_{+}
\end{pmatrix}
\end{equation}
To compute the Liouville speed limit, the matrix $\mathcal{H}-\mathcal{H}^{\dagger}$ is required:
\begin{eqnarray}
\mathcal{H}-\mathcal{H}^{\dagger} &=&
ib_{+}\begin{pmatrix}
\mathbf{N}_{5\times 5} & \mathbf{e}_{5\times 4}(1,4)   \\
\mathbf{e}_{4\times 1}(1,4)  & \mathbf{N}'_{4\times 4}
\end{pmatrix} + \nonumber \\
&&
ic_{+}\begin{pmatrix}
\mathbf{0}_{5\times 5} &  \mathbf{e}_{5\times 4}(5,4)  \\
 \mathbf{e}_{4\times 5}(4,5)  & \mathbf{0}_{4\times 4}
\end{pmatrix} \nonumber
\end{eqnarray}
where
\begin{small}
\begin{equation}
\mathbf{N}_{5\times 5}=
\begin{pmatrix}
-i(\gamma_{31}+\gamma_{21}) & 0 & 0  &0& ia_{+} \\
0 & -i2\Gamma &0 &0 &0 \\
0 & 0 & -i2\Gamma &0 &0 \\
0 & 0 &0 & -i2\Gamma &0 \\
ia_{+} & 0 &0 &0 & -i2(\gamma_{12}+\gamma_{32})
\end{pmatrix}
\end{equation}
\end{small}
and
\begin{equation}
\mathbf{N}'_{4\times 4}=
\begin{pmatrix}
-i2\Gamma & 0 & 0  &0 \\
0 & -i2\Gamma &0 &0  \\
0 & 0 & -i2\Gamma &0 \\
0 & 0 &0 & -i2(\gamma_{13}+\gamma_{23})  \\
\end{pmatrix}
\end{equation}
$||\mathcal{H}-\mathcal{H}^{\dagger} ||_{sp}$ is the absolute value of the greatest zero of its characteristic polynomial $(A_3 z^3 + A_2 z^2 + A_1 z + A_0)(z+i2\Gamma)^6$ with $A_3=1/2$, $A_2=i(a_{+}+b_{+}+c_{+})$,
\begin{eqnarray}
A_1 & =&  \tfrac{1}{2}\gamma_{12}^2 -\gamma_{12}(2\gamma_{13}+\gamma_{21}+2\gamma_{23}+2\gamma_{31}) + \nonumber \\
& & +\tfrac{1}{2}\gamma_{13}^2-(2\gamma_{21}+\gamma_{31}+2\gamma_{32})\gamma_{13} + \tfrac{1}{2}\gamma_{21}^2 -2c_{+}\gamma_{21}+ \nonumber \\
& &\tfrac{1}{2}\gamma_{23}^2-(2\gamma_{31}+\gamma_{32})\gamma_{23} ++\tfrac{1}{2}\gamma_{31}^2 - 2\gamma_{31}\gamma_{32}+\tfrac{1}{2}\gamma_{32}^2, \nonumber
\end{eqnarray}
and
\begin{eqnarray}
A_0 &=& i(\gamma_{13}+\gamma_{23})\gamma_{12}^2  + ( i \gamma_{13}^2 + i( -2\gamma_{21} + \gamma_{23}-2\gamma_{32}+ \nonumber\\
& & +\gamma_{32})\gamma_{13} -2i\gamma_{21}\gamma_{23} - 3i(\gamma_{23}-\tfrac{1}{3}\gamma_{31}-\tfrac{1}{3}\gamma_{32})\gamma_{31})\gamma_{12} \nonumber\\
& & + i\gamma_{13}^2\gamma_{32} + (i\gamma_{21}^2+i(\gamma_{23}-3\gamma_{32})\gamma_{21}-2i\gamma_{31}\gamma_{32})\gamma_{13} \nonumber \\
&& i(\gamma_{21}+\gamma_{31})(\gamma_{21}\gamma_{23}+\gamma_{23}^2-2\gamma_{23}\gamma_{32}+\gamma_{32}(\gamma_{31}+\gamma_{32})). \nonumber
\end{eqnarray}
The computation of the Hilbert speed limit requires the determination of $4||\mathbf{h}||_{1}=4\sum_{l,m}^{N^2-1}|a_{ml}|$. In this case, this term can be expressed as:
\begin{eqnarray}
||\mathbf{h}||_{1} &=&  |\Gamma-\tfrac{1}{6}a_{+}-\tfrac{2}{3}\gamma_{13}-\tfrac{2}{3}\gamma_{23}-\tfrac{1}{6}\gamma_{31}-\tfrac{1}{6}\gamma_{32}| \nonumber \\
&& + \tfrac{\sqrt{3}}{3}|a_{-}-\gamma_{31}+\gamma_{32}| +b_{+}+|b_{-}| + a_{+} + |a_{-}| \nonumber\\
&& |\Gamma-\tfrac{1}{2}a_{+}-\tfrac{1}{2}\gamma_{31}-\tfrac{1}{2}\gamma_{32}|+c_{+} + |c_{-}|  \nonumber
\end{eqnarray}
We consider the numerical example of Sec.~\ref{sec4} with $\gamma_{12}=1$, $\gamma_{13}=1/2$, $\gamma_{23}=1/2$, $\gamma_{31}=0$, $\gamma_{21}=0$ and $\gamma_{32}=0$ thus, the two lower bounds are:
\begin{eqnarray}
\Delta t_{L} &=& \frac{\ln(3)}{\max\left(2\Gamma,1+\tfrac{\sqrt{10}}{2} \right)} \\
\Delta t_{H} &=& \frac{\ln(3)}{16+4\tfrac{\sqrt{3}}{3} + 4|\Gamma-5/6|+4|\Gamma-1/2|}
\end{eqnarray}
If the dephasing rate $\Gamma$ goes to infinity, namely $\Gamma\gg 1$, then:
\begin{eqnarray}
\Delta t_{L} &\simeq& \frac{\ln(3)}{2\Gamma} \\
\Delta t_{H} &\simeq& \frac{\ln(3)}{2^3\Gamma}
\end{eqnarray}
where the initial state is $\rho(0)=\mathbf{e}_{3\times 3}(1,1)$ and the final one is the maximally mixed state given by $\rho(t_f)=\textrm{diag}(\tfrac{1}{3},\tfrac{1}{3},\tfrac{1}{3})$.
\section{Dynamics of a dissipative three-level quantum system}\label{secappb}
We derive in this paragraph the differential equations governing the dynamics of a dissipative three-level quantum system in the coherence vector formalism. For a general density matrix $\rho$ of the form:
$$
\rho=\begin{pmatrix}\rho_{11} & \rho_{12} & \rho_{13} \\
\rho_{21} & \rho_{22} & \rho_{23} \\
\rho_{31} & \rho_{32} & \rho_{33}
\end{pmatrix}
$$
we have:
$$
\begin{cases}
s_0=\frac{1}{\sqrt{3}} \\
s_1=\frac{1}{\sqrt{2}}(\rho_{12}+\rho_{21}); s_2=\frac{i}{\sqrt{2}}(\rho_{12}-\rho_{21}) \\
s_3=\frac{1}{\sqrt{2}}(\rho_{13}+\rho_{31}); s_4=\frac{i}{\sqrt{2}}(\rho_{13}-\rho_{31}) \\
s_5=\frac{1}{\sqrt{2}}(\rho_{23}+\rho_{32}); s_6=\frac{i}{\sqrt{2}}(\rho_{23}-\rho_{32}) \\
s_7=\frac{1}{\sqrt{2}}(\rho_{11}-\rho_{22}); s_8=\frac{1}{\sqrt{6}}(\rho_{11}+\rho_{22}-2\rho_{33})
\end{cases}
$$
If the unitary dynamics of the density matrix are generated by:
$$
H_I=
\begin{pmatrix}
0 & u & 0 \\
u^* & 0 & v \\
0 & v^* & 0
\end{pmatrix}
$$
where the control fields are expressed as $u=u_1+iu_2$ and $v=v_1+iv_2$, it can be shown that the coordinates of the coherence vector fulfill the differential system:
$$
\begin{cases}
\dot{s}_1=-2u_2s_7+v_1s_4+v_2s_3-\Gamma s_1 \\
\dot{s}_2=-2u_1s_7-v_1s_3+v_2s_4-\Gamma s_2 \\
\dot{s}_3=u_2s_5-u_1s_6-v_2s_1+v_1s_2-\Gamma s_3 \\
\dot{s}_4=-v_1s_1-v_2s_2+u_1s_5+u_2s_6-\Gamma s_4 \\
\dot{s}_5=-u_1s_4-u_2s_3+v_2(-\sqrt{3}s_8+s_7)-\Gamma s_5 \\
\dot{s}_6=u_1s_3-u_2s_4+v_1(s_7-\sqrt{3}s_8)-\Gamma s_6 \\
\dot{s}_7=2u_1s_2+2u_2s_1-v_1s_6-v_2s_5+L(s_7) \\
\dot{s}_8=\sqrt{3}v_1s_6+\sqrt{3}v_2s_5+L(s_8)
\end{cases}
$$
with
$$
\begin{cases}
L(s_7)=\frac{1}{3\sqrt{2}}[-2\gamma_{21}-\gamma_{31}+2\gamma_{12}+\gamma_{32}+\gamma_{13}-\gamma_{23}]\\
+\frac{s_7}{2}[-2\gamma_{21}-\gamma_{31}-2\gamma_{12}-\gamma_{32}]\\
+\frac{s_8}{2\sqrt{3}}[-2\gamma_{21}-\gamma_{31}+2\gamma_{12}+\gamma_{32}-2\gamma_{13}+2\gamma_{23}] \\
L(s_8)=\frac{1}{\sqrt{6}}[-\gamma_{31}-\gamma_{32}+\gamma_{13}+\gamma_{23}]\\
+\frac{\sqrt{3}}{2}s_7[-\gamma_{31}+\gamma_{32}]\\
+\frac{s_8}{2}[-\gamma_{31}-\gamma_{32}-2\gamma_{13}-2\gamma_{23}]
\end{cases}
$$
With the notations of Sec.~\ref{sec4}, we have:
$$
\begin{cases}
L(s_7)=q_7+r_{77}s_7+r_{78}s_8 \\
L(s_8)=q_8+r_{87}s_7+r_{88}s_8
\end{cases}
$$
\section{Time evolution of the Lagrange multiplier $\mu$}\label{secappc}
We describe in this paragraph the computation of the time evolution of $\mu$ in the magic subspace $\mathcal{M}_d$ for two- and three- level quantum systems. In each case, the final goal is to compute $t_d$ the time to go from $\mathcal{M}_o$ to the zero coherence vector.

We first consider the two-level quantum system analyzed in Sec.~\ref{sec3}. The purity $p_d=s_3^2$ in $\mathcal{M}_d$ is governed by the following differential equation:
$$
\dot{p}_d=2\gamma_-s_3-2\gamma_+s_3^2.
$$
Using the relation $p_d=\frac{\gamma_-^2}{4(\mu-\gamma_+)^2}$, we deduce that the dynamics of $\mu$ are given by:
$$
\dot{\mu}=(\mu-\gamma_+)(2\mu-\gamma_+),
$$
which leads to:
\begin{equation}\label{eqmu2}
\mu(t)=\frac{\gamma_+(2\Gamma-\gamma_+)-\gamma_+(\Gamma-\gamma_+)e^{\gamma_+t}}{(2\Gamma-\gamma_+)-2(\Gamma-\gamma_+)e^{\gamma_+t}}.
\end{equation}
with $\mu(0)=\Gamma$. The zero coherence vector is reached when $\mu\to +\infty$, i.e. when the denominator of Eq.~\eqref{eqmu2} is zero. Finally, we arrive at:
$$
t_d=\frac{1}{\gamma_+}\ln [\frac{2\Gamma-\gamma_+}{2(\Gamma-\gamma_+)}],
$$
which is the control time used in Sec.~\ref{sec3}.

For the three-level quantum system described in Sec.~\ref{sec4}, $s_7$ and $s_8$ are solutions of the following system:
$$
\begin{cases}
q_7+2r_{77}s_7+(r_{78}+r_{87})s_8+2\mu s_7 =0\\
q_8+2r_{88}s_8+(r_{78}+r_{87})s_7+2\mu s_8=0,
\end{cases}
$$
which leads to:
\begin{equation}\label{eqs78}
\begin{cases}
s_7 = [q_8(r_{78}+r_{87})-2q_7(r_{88}+\mu)]/D \\
s_8 = [q_7(r_{78}+r_{87})-2q_8(r_{77}+\mu)]/D ,
\end{cases}
\end{equation}
where $D=4(r_{77}+\mu)(r_{88}+\mu)-(r_{78}+r_{87})^2$. Starting from the relation $p_d=s_7^2+s_8^2$, we can derive the differential equation verified by $\mu(t)$. First, we have:
$$
\dot{p}_d=2s_7\dot{s}_7+2s_8\dot{s}_8=2(s_7L(s_7)+s_8L(s_8)).
$$
This time derivative can also be expressed as:
$$
\dot{p}_d=2(s_7\frac{d s_7}{d\mu}+s_8\frac{d s_8}{d\mu})\dot{\mu}.
$$
Identifying the two expressions of $\dot{p}_d$, we arrive after straightforward computations at:
\begin{equation}\label{eqmut}
\dot{\mu}=\frac{q_7s_7+q_8s_8-2\mu (s_7^2+s_8^2)}{2(s_7\frac{d s_7}{d\mu}+s_8\frac{d s_8}{d\mu})}.
\end{equation}
Using Eq.~\eqref{eqs78}, this differential equation allows us to compute numerically the time evolution of $\mu$.


\begin{thebibliography}{99}

\bibitem{glaser15} S. J. Glaser, U. Boscain, T. Calarco, C. P. Koch, W. K\"ockenberger,
R. Kosloff, I. Kuprov, B. Luy, S. Schirmer, T. Schulte-Herbr\"uggen, D. Sugny, and F. K. Wilhelm, Eur. Phys. J. D \textbf{69}, 279 (2015)

\bibitem{brif} C. Brif, R. Chakrabarti, and H. Rabitz, New J. Phys. \textbf{12}, 075008 (2010)

\bibitem{RMP} C. P. Koch, M. Lemeshko and D. Sugny, Rev. Mod. Phys. \textbf{91}, 035005 (2019)

\bibitem{dong} D. Dong and I. A. Petersen, IET Control Theory A \textbf{4}, 2651 (2010)

\bibitem{alessandrobook} D. D’Alessandro, \emph{Introduction to Quantum Control and Dynamics} (Chapman and Hall, Boca Raton, FL, 2008)

\bibitem{pont} L. S. Pontryagin et al., \emph{The Mathematical Theory of Optimal Processes} (John Wiley and Sons, New York, 1962).

\bibitem{alessandro} D. D’Alessandro, IEEE Trans. Autom. Control \textbf{46}, 866 (2001)

\bibitem{boscain} U. Boscain and P. Mason, J. Math. Phys. \textbf{47}, 062101 (2006)

\bibitem{hegerfeldt} G. C. Hegerfeldt, Phys. Rev. Lett. \textbf{111}, 260501 (2013)

\bibitem{garon} A. Garon, S. J. Glaser, and D. Sugny, Phys. Rev. A \textbf{88}, 043422
(2013)

\bibitem{khaneja1} N. Khaneja, R. Brockett, and S. J. Glaser, Phys. Rev. A \textbf{63}, 032308 (2001)

\bibitem{khaneja2} N. Khaneja, S. J. Glaser, and R. Brockett, Phys. Rev. A \textbf{65}, 032301 (2002)

\bibitem{lapert} M. Lapert, Y. Zhang, M. Braun, S. J. Glaser, and D. Sugny, Phys. Rev. Lett. \textbf{104}, 083001 (2010)

\bibitem{bonnard} B. Bonnard, O. Cots, S. J. Glaser, M. Lapert, D. Sugny, and Y. Zhang, IEEE Trans. Automat. Control \textbf{57}, 1957 (2012)

\bibitem{grape} N. Khaneja, T. Reiss, C. Kehlet, T. Schulte-Herbrüggen, and S. J. Glaser, J. Magn. Reson. \textbf{172}, 296 (2005)

\bibitem{reich} D. M. Reich, M. Ndong, and C. P. Koch, J. Chem. Phys. \textbf{136}, 104103 (2012)

\bibitem{gross} J. Werschnik and E. K. U. Gross, J. Phys. B \textbf{40}, R175 (2007)

\bibitem{doria} P. Doria, T. Calarco, and S. Montangero, Phys. Rev. Lett. \textbf{106}, 190501 (2011)

\bibitem{calarco} T. Caneva, M. Murphy, T. Calarco, R. Fazio, S. Montangero, V. Giovannetti, and G. E. Santoro, Phys. Rev. Lett. \textbf{103}, 240501 (2009)

\bibitem{QSLreview} S. Deffner and S. Campbell, J. Phys. A: Math. Theor. \textbf{50}, 453001 (2017)

\bibitem{QSLreview2} M. R. Frey, Quantum Inf. Process., \textbf{15}, 3919 (2016)

\bibitem{lloyd} S. Lloyd, Nature \textbf{406}, 1047 (2000)

\bibitem{giovannetti} V. Giovannetti, S. Lloyd, and L. Maccone, Phys. Rev. A \textbf{67}, 1 (2003)

\bibitem{alipour} S. Alipour, M. Mehboudi, and A. T. Rezakhani, Phys. Rev. Lett. \textbf{112}, 120405 (2014)

\bibitem{giovannetti2} V. Giovannetti, S. Lloyd, and L. Maccone, Nat. Photonics \textbf{5}, 222 (2011)

\bibitem{chin} A. W. Chin, S. F. Huelga, and M. B. Plenio, Phys. Rev. Lett. \textbf{109}, 233601 (2012)

\bibitem{demkowicz} R. Demkowicz-Dobrza\'nski, J. Kolody\'nski, and M. Guta, Nat. Commun. \textbf{3}, 1063 (2012)

\bibitem{deffnerthermo} S. Deffner, Phys. Rev. Research \textbf{2}, 013161 (2020)

\bibitem{qthermo} F. Campaioli, F. A. Pollock, F. C. Binder, L. C\'eleri, J. Goold, S. Vinjanampathy, and K. Modi, Phys. Rev. Lett. \textbf{118}, 150601 (2017)

\bibitem{shanadan} B. Shanahan, A. Chenu, N. Margolus and A. del Campo, Phys. Rev. Lett. \textbf{120}, 070401 (2018)

\bibitem{okuyama} M. Okuyama and M. Ohzeki, Phys. Rev. Lett. \textbf{120}, 070402 (2018)

\bibitem{pires} D. P. Pires, M. Cianciaruso, L. C. C\'eleri, G. Adesso, and D. O. Soares-Pinto, Phys. Rev. X \textbf{6}, 021031 (2016)

\bibitem{campaioli} F. Campaioli, F. A. Pollock, F. C. Binder, and K. Modi, Phys. Rev. Lett. \textbf{120}, 060409 (2018)

\bibitem{bason} M. G. Bason, M. Viteau, N. Malossi, P. Huillery, E. Arimondo, D. Ciampini, R. Fazio, V. Giovannetti, R. Mannella, and O. Morsch. Nat. Phys. \textbf{8}, 147 (2012)

\bibitem{campaioli2} F. Campaioli, F. A. Pollock, and K. Modi, Quantum \textbf{3}, 168 (2019)

\bibitem{lidar} I. Marvian and D. A. Lidar, Phys. Rev. Lett. \textbf{115}, 210402 (2015)

\bibitem{deffner} S. Deffner and E. Lutz, Phys. Rev. Lett. \textbf{111}, 010402 (2013)

\bibitem{taddei} M. M. Taddei, B. M. Escher, L. Davidovich, and R. L. de Matos Filho, Phys. Rev. Lett. \textbf{110}, 050402 (2013)

\bibitem{campo} A. del Campo, I. L. Egusquiza, M. B. Plenio, and S. F. Huelga, Phys. Rev. Lett. \textbf{110}, 050403 (2013)

\bibitem{kosloff} R. Uzdin and R. Kosloff, Eur. Phys. Lett. \textbf{115}, 40003 (2016)

\bibitem{brody} D. C. Brody and B. Longstaff, Phys. Rev. Research \textbf{1}, 033127 (2019)

\bibitem{funo} K. Funo, N. Shiraishi and K. Saito, New J. Phys. \textbf{21}, 013006 (2019)

\bibitem{hutter} A. Hutter and S. Wehner, Phys. Rev. Lett. \textbf{108}, 070501 (2012)

\bibitem{altafini1} C. Altafini, Phys. Rev. A \textbf{70}, 062321 (2004)

\bibitem{altafini2} C. Altafini, J. Math. Phys. \textbf{44}, 2357 (2003)

\bibitem{dive} B. Dive, D. Burgarth and F. Mintert, Phys. Rev. A \textbf{94}, 012119 (2016)

\bibitem{schulte1} G. Dirr, U. Helmke, I. Kurniawan and T. Schulte-Herbrueggen, Rep. Math. Phys. \textbf{64}, 93 (2009)

\bibitem{schulte2} F. Von Ende, G. Dirr, M. Keyl and T. Schulte-Herbrueggen, Open systems and Information dynamics \textbf{26}, 1950014 (2019)

\bibitem{kochreview} C. P. Koch, J. Phys.: Condens. Matter \textbf{28}, 213001 (2016)

\bibitem{lapert:2013} M. Lapert, E. Ass\'emat, S. J. Glaser and D. Sugny, Phys. Rev. A \textbf{88}, 033407 (2013)

\bibitem{mukherjee} V. Mukherjee, A. Carlini, A. Mari, T. Caneva, S. Montangero, T. Calarco, R. Fazio, and
V. Giovannetti, Phys. Rev. A \textbf{88}, 062326 (2013)

\bibitem{tannor} D. J. Tannor and A. Bartana, J. Phys. Chem. A \textbf{103}, 10359 (1999)

\bibitem{bonnard1} B. Bonnard and D. Sugny, SIAM J. Control Optim. \textbf{48}, 1289 (2009)

\bibitem{bonnard2} B. Bonnard, M. Chyba, and D. Sugny, IEEE Trans. Autom. Control \textbf{54}, 2598 (2009)

\bibitem{khaneja3} S. E. Sklarz, D. J. Tannor, and N. Khaneja, Phys. Rev. A \textbf{69}, 053408 (2004)

\bibitem{schirmer} S. G. Schirmer, T. Zhang and J. V. Leahy, J. Phys. A \textbf{37}, 1389 (2004)

\bibitem{DFS} D. A. Lidar, I. L. Chuang and K. B. Whaley, Phys. Rev. Lett. \textbf{81}, 2594 (1998)

\bibitem{breuerbook} H.-P. Breuer and F. Petruccione, \emph{The theory of open quantum systems} (Oxford University, Oxford, 2002)

\bibitem{lindblad} G. Lindbald, Commun. Math. Phys. \textbf{48}, 119 (1976)

\bibitem{gorini} V. Gorini, A. Kossakowski, and E. C. G. Sudarshan, J. Math.Phys. \textbf{17}, 821 (1976)

\bibitem{schirmer01} S. G. Schirmer, H. Fu, and A. I. Solomon, Phys. Rev. A \textbf{63}, 063410 (2001)

\bibitem{alickibook} R. Alicki and K. Lendi, \emph{Quantum Dynamical Semigroups and Application} (Springer, Berlin, 1987).

\bibitem{schirmer:04} S. G. Schirmer and A. I. Solomon, Phys. Rev. A \textbf{70}, 022107 (2004)




\end{thebibliography}
\end{document}